\documentclass[aps,prb,twocolumn,superscriptaddress,floatfix,amsmath]{revtex4}

\usepackage{graphicx}
\usepackage{dcolumn}
\usepackage{bm}
\usepackage{natbib}
\usepackage{color}
\usepackage{amsmath}
\usepackage{amssymb}

\renewcommand{\vec}[1]{\boldsymbol{#1}}

\begin{document}

\title{Low temperature specific heat and possible gap to magnetic excitations in  
       the Heisenberg pyrochlore antiferromagnet Gd$_2$Sn$_2$O$_7$}

\author{Adrian Del Maestro}
\affiliation{Department of Physics, Harvard University, Cambridge, Massachusetts, 02138, USA}

\author{Michel J. P. Gingras}
\affiliation{Department of Physics and Astronomy, University of Waterloo, Waterloo,
Ontario, N2L 3G1, Canada}
\affiliation{Department of Physics and Astronomy, University of Canterbury, 
Private Bag 4800, Christchurch, New Zealand}

\date{\today}

\begin{abstract}
The Gd$_2$Sn$_2$O$_7$ pyrochlore Heisenberg antiferromagnet displays a
phase transition to a four sublattice N\'eel ordered state at a 
critical temperature 
$T_c \sim 1$~K.  The low-temperature state found via neutron scattering corresponds to
that predicted by a classical model that considers nearest-neighbor
antiferromagnetic exchange and long-range dipolar interactions.
Despite the seemingly conventional nature of the ordered state,
the specific heat $C_v$ has been found to be described 
in the temperature range $350~\mathrm{mK} \le T \le 800~\mathrm{mK}$ 
by an anomalous power law, $C_v\sim T^2$.
A similar temperature dependence of $C_v$ has also been reported for
Gd$_2$Ti$_2$O$_7$, another pyrochlore Heisenberg material.
Such behavior is 
to be contrasted with the typical $T^3$ behavior expected for 
a three-dimensional antiferromagnet with conventional long-range order which is
then generally accompanied by an $\exp(-\Delta/T)$ behavior at lower
temperature where anisotropy effects induce
 a gap $\Delta$ to collective spin excitations.
Such anomalous $T^2$ behavior in $C_v$ 
has been argued to be correlated to an unusual 
energy-dependence of the density of states which also seemingly manifests itself in
low-temperature spin fluctuations found in muon spin relaxation experiments.
In this paper, we report calculations of $C_v$ that consider spin wave
like excitations out of the  N\'eel order observed in Gd$_2$Sn$_2$O$_7$
via neutron scattering.
We argue that the parametric $C_v \propto T^2$ does not reflect the true
low-energy excitations of Gd$_2$Sn$_2$O$_7$.
Rather, we find that the low-energy excitations of 
this material are antiferromagnetic magnons gapped by single-ion and dipolar anisotropy
effects, and that the lowest temperature of 350~mK considered in previous
specific heat measurements accidentally happens to
coincide with a crossover temperature 
{\it below} which magnons become thermally activated and $C_v$ takes
an exponential form.
We argue that further specific heat measurements that extend down to at least
100~mK are required in order to ascribe an unconventional description of
magnetic excitations out
of the ground state of Gd$_2$Sn$_2$O$_7$ 
or to invalidate the standard picture of gapped excitations proposed herein.

\end{abstract}


\maketitle


\section{Introduction}
\label{sec:Introductino}

\subsection{Persistent spin dynamics in pyrochlores}

A magnetic system with  Heisenberg spins that sit on the vertices of 
a three-dimensional pyrochlore lattice of corner sharing tetrahedra 
and interact among themselves via nearest-neighbor antiferromagnetic exchange interactions 
is highly geometrically frustrated \cite{Reimers-MFT,Moessner:Pyro,Moessner-CJP}.
Such a system is theoretically predicted to not develop conventional magnetic long range order at 
finite temperature for either classical \cite{Moessner-CJP,Moessner-PRB}
or quantum spins \cite{Canals:Pyro}.
As a result of this frustration, 
real magnetic materials with antiferromagnetically coupled spins on
this pyrochlore structure are highly sensitive
to weak perturbative interactions beyond  nearest-neighbor exchange which dramatically
affect the nature of the low temperature state.
It is partially for this reason that the 
insulating R$_2$M$_2$O$_7$ magnetic pyrochlore oxides 
have attracted such a great deal of attention in recent years \cite{Greedan:Review}.
Indeed, this family of materials has been found to display a variety of magnetic states 
and exotic low temperature behaviors that strongly
depend on the specific elements R and M considered \cite{Greedan:Review}.

In  R$_2$M$_2$O$_7$, the R site is occupied by a trivalent ion, such as 
diamagnetic Y$^{3+}$  or a magnetic rare earth ion
(R=Gd$^{3+}$, 
Tb$^{3+}$, 
Dy$^{3+}$, 
Ho$^{3+}$, 
Er$^{3+}$, 
Tm$^{3+}$, 
Yb$^{3+}$) 
while the M site is occupied by a tetravalent ion 
which can be diamagnetic, such as Ti$^{4+}$ or Sn$^{4+}$, 
or magnetic, such as Mo$^{4+}$ or Mn$^{4+}$.
Both the R and M sites form distinct interpenetrating
lattices of corner-shared tetrahedra and 
either site can be magnetic or non-magnetic\cite{Reimers-MFT,Greedan:Review}.
	Such freedom allows for a large variety of phenomenology in the pyrochlore materials.  This
	includes spin glasses like Y$_2$Mo$_2$O$_7$
	\cite{Greedan:SG-YMO,Gingras:YMO,Gardner-YMO,Keren:YMO}
	and Tb$_2$Mo$_2$O$_7$ \cite{Gaulin-Tb2Mo2O7}, spin ices such as 
	Ho$_2$Ti$_2$O$_7$ \cite{Harris-PRL,Bramwell-PRL,Cornelius}
	and Dy$_2$Ti$_2$O$_7$ \cite{Ramirez-Nature}, conventional long range ordered materials like
	Gd$_2$Ti$_2$O$_7$ \cite{Raju:GTO,Ramirez-GTO,Stewart:GTO,Champion:ETO} 
and Gd$_2$Sn$_2$O$_7$ \cite{Wills:GSO} and 
	even possible spin liquids as in the case of 
Tb$_2$Ti$_2$O$_7$ 
\cite{Gardner-PRL,Gardner-PRB-50mk,Kao:TTO,Enjalran-review,Mirebeau-review,Molavian}.

One common thread throughout these various materials is that 
several experimental studies have found that,
almost without exceptions \cite{Bonville:GSO1,Gardner-dynamics-review},
{\it all} insulating rare-earth pyrochlore materials 
R$_2$Ti$_2$O$_7$ and R$_2$Sn$_2$O$_7$ display 
temperature-independent spin dynamics down 
to $T_0 \sim O(10^1)~{\rm mK}$.  
Indeed, residual low temperature dynamics has been found in 
pyrochlore magnetic
materials with low temperature states that range from not 
understood whatsoever \cite{Gardner-PRL,YTO-muSR} to seemingly
conventional long range ordered \cite{Raju:GTO,Ramirez-GTO,Wills:GSO,Stewart:GTO,Champion:ETO}.
We note in passing that persistent low-temperature spin dynamics has also
been found in the Gd$_3$Ga$_5$O$_{12}$ garnet (GGG) \cite{Dunsiger:GGG,Marshall:GGG}
and in the SrCr$_8$Ga$_4$O$_{19}$ kagome antiferromagnet \cite{Uemura:SCGO}.
We now briefly review the various experimentally
observed behaviors of the R$_2$Ti$_2$O$_7$ and R$_2$Sn$_2$O$_7$ pyrochlore oxides.


Strong evidence for fluctuating spins down to extremely low temperatures
has been observed in Tb$_2$Ti$_2$O$_7$ where Ti$^{4+}$ 
at the M site is non-magnetic \cite{Gardner-PRL}.
The reason for the failure of Tb$_2$Ti$_2$O$_7$ to develop long-range magnetic order
above 50~mK \cite{Gardner-PRB-50mk} despite a Curie-Weiss temperature, 
$\theta_{\rm CW} \sim -14~{\rm K}$ remains to this day largely
unexplained \cite{Enjalran-review,Mirebeau-review,Molavian}.
Yb$_2$Ti$_2$O$_7$ is perhaps just as intriguing, with 
specific heat measurements 
revealing a sharp first order transition
at $T_c \approx 0.24$ K \cite{YTO-muSR,YTO-neutrons}, but with 
the spins not appearing static below $T_c$ since 
muon spin relaxation ($\mu$SR) and M\"ossbauer spectroscopy
finds significant spin dynamics down to the lowest temperature \cite{YTO-muSR}.
Hence, the observed first order transition in Yb$_2$Ti$_2$O$_7$ seems rather 
unconventional.

Ho$_2$Ti$_2$O$_7$ \cite{Harris-PRL,Bramwell-PRL,Cornelius}
and Dy$_2$Ti$_2$O$_7$ \cite{Ramirez-Nature}
are frustrated ferromagnets \cite{Harris-PRL} and possess an extensive low-temperature 
magnetic entropy \cite{Cornelius,Ramirez-Nature} similar to that of
the common hexagonal 
I$_{\rm h}$ phase of 
water ice \cite{Pauling,Giauque}. 
As such, the (Ho,Dy)$_2$(Ti,Sn)$_2$O$_7$ materials are referred to as 
{\it spin ices} \cite{Bramwell-Science}.
Theoretical and numerical studies have  shown that the spin ice behavior originates from the
long range nature of magnetic dipole-dipole interactions \cite{Bramwell-Science,Hertog-PRL}.
Numerical Monte Carlo studies using non-local loop dynamics predict that those interactions should 
lead to long range order at low temperatures \cite{Melko:JPC}.
Yet, at variance with the numerical predictions, experimental studies of the Dy$_2$Ti$_2$O$_7$ 
\cite{Fukuyama-DTO}  and Ho$_2$Ti$_2$O$_7$ \cite{Harris-PRL,Harris-JMMM,Ehlers}  have not found a
transition to long range order down to 60~mK.
In particular, muon spin relaxation ($\mu$SR)~\cite{Harris-JMMM} and neutron spin echo 
\cite{Ehlers} experiments
find evidence for Ho$^{3+}$ spin dynamics well below $T_{SI}$ in Ho$_2$Ti$_2$O$_7$.
Interestingly, a recent neutron scattering study on Tb$_2$Sn$_2$O$_7$ 
found a transition to a long-range ordered state at $T_c\approx 0.87~{\rm K}$, with
an analysis of the scattering intensity indicating that the observed state is a long-range 
spin ice state \cite{Mirebeau-TSO}.
However, even more recent $\mu$SR studies find that the state at $T<T_c$ in 
Tb$_2$Sn$_2$O$_7$ remains dynamic down to the lowest temperature \cite{Yaouanc-TSO,Bert-TSO}.
Er$_2$Ti$_2$O$_7$, like Tb$_2$Sn$_2$O$_7$, 
was found via neutron scattering to display long range order below 1.2 K ~\cite{Champion:ETO}.
Yet, $\mu$SR found persistent spin dynamics down to the lowest temperature ~\cite{Lago:ETO}.

Gd$_2$Ti$_2$O$_7$ displays two consecutive transitions at $T_c^+\sim$ 1 K and  $T_c^- \sim 0.7~{\rm K}$
 \cite{Raju:GTO,Ramirez-GTO}. 
 Neutron scattering experiments~\cite{Stewart:GTO} found that the magnetic state
between $T_c^-$ and $T_c^+$ has one site out of four on a tetrahedral unit cell that is
paramagnetic and fluctuating. At $T<T_c^-$, the fourth site orders, but remains much more
dynamic than the three other sites. The microscopic mechanism giving 
rise to the two experimentally observed states is still not understood.
Here too, in Gd$_2$Ti$_2$O$_7$, $\mu$SR finds considerable spin dynamics persisting
down to 20~mK \cite{Yaouanc-GTO-PRL,Dunsiger:GTO}. 
A phenomenological model for the density of states, $g(\epsilon)$,
has been proposed for the low-temperature state
of Gd$_2$Ti$_2$O$_7$. Most significantly, 
the proposed model for $g(\epsilon)$ was shown to describe the 
peculiar temperature dependence of the magnetic specific heat, $C_v(T)$,
in Gd$_2$Ti$_2$O$_7$ which was found to be $C_v(T) \propto T^2$ below $T_c^-$. 
Such $C_v\propto T^2$ behavior is rather unconventional.
Indeed, in a conventional long-range ordered three-dimensional antiferromagnet, 
$C_v \sim T^3$ down to a temperature where the temperature dependence turns to
$C_v \sim \exp(-\Delta/T)$ because of a gap $\Delta$ in the excitation spectrum
induced by single ion anisotropy or anisotropic spin-spin interactions.


In all the pyrochlore systems reviewed above,
Tb$_2$Ti$_2$O$_7$, 
Yb$_2$Ti$_2$O$_7$, 
(Ho,Dy)$_2$Ti$_2$O$_7$,  
Tb$_2$Sn$_2$O$_7$,
Er$_2$Ti$_2$O$_7$,
and Gd$_2$Ti$_2$O$_7$, 
some theoretical lapses exist in our understanding of the {\it equilibrium} 
thermodynamic low-temperature state. Hence, it is  perhaps not completely
surprising that the spin dynamics appears unconventional in these materials 
with, in particular, a temperature independent $\mu$SR spin 
polarization relaxation rate  down to a baseline temperature 
$T_0\sim 10^1~{\rm mK}$.  However, that tentative self-reassured standpoint 
is put on shaky ground by the $\mu$SR and specific heat measurements on Gd$_2$Sn$_2$O$_7$ 
that we now discuss.


\subsection{The case of Gd$_2$Sn$_2$O$_7$}

It was first proposed that the aforementioned Gd$_2$Ti$_2$O$_7$ 
material would be a good candidate for a 
classical Heisenberg pyrochlore antiferromagnet with leading perturbations coming from 
long-range magnetic dipole-dipole interactions \cite{Raju:GTO}. The reason for this is that
Gd$^{3+}$ is an S-state ion with half-filled 4f shell, hence orbital angular moment $\mathrm{L}=0$,
and spin $\mathrm{S}=7/2$. Spin anisotropy is therefore expected to be much smaller than for
the above Ho, Dy and Tb based rare earth materials \cite{Jensen:REM}.
In that context, Gd$_2$Sn$_2$O$_7$ should be similar to Gd$_2$Ti$_2$O$_7$;
the main difference being that 
Gd$_2$Sn$_2$O$_7$ displays 
only one phase transition
observed from a paramagnetic to a long-range ordered phase at $T_c\sim 1$~K ~\cite{Wills:GSO}.  
Perhaps most interestingly, unlike Gd$_2$Ti$_2$O$_7$, 
the experimentally observed long-range 
ordered phase in Gd$_2$Sn$_2$O$_7$ corresponds to the one predicted by Palmer and Chalker for the
classical Heisenberg pyrochlore antiferromagnet 
model with perturbative long-range dipolar interactions \cite{PC}.
It is possible that the experimentally observed transition in Gd$_2$Sn$_2$O$_7$ corresponds to two
very close transitions \cite{Enjalran:GTO,Cepas:MFT} that are not resolved \cite{Cepas:MC}.

From our perspective, Gd$_2$Sn$_2$O$_7$ is an exemplar of the intriguing behavior discussed above.
Yet, it offers itself as a crucial system to understand. The reasons are as follows:
(i) as in Gd$_2$Ti$_2$O$_7$, and all the materials previously described,
persistent low-temperature spin dynamics have been observed \cite{Bonville:GSO2} and (ii)
again similarly to Gd$_2$Ti$_2$O$_7$, unconventional power-law temperature dependence of the magnetic
specific heat has been found, specifically, $C_v \sim T^2$. 
So here too, there may exists the possibility to relate a dynamical response and a bulk thermodynamic
measurements to an unconventional density of states $g(\epsilon)$ $-$ a possible manifestation of
the spectral down-shift that corresponds to the hallmark of highly frustrated systems.
We see the experimental results on Gd$_2$Sn$_2$O$_7$ as a crucial paradox to contend with.
Since the observed ordered state in Gd$_2$Sn$_2$O$_7$ corresponds to the one 
predicted by the model of Palmer and  Chalker \cite{PC}, or a more
refined model that includes Gd$^{3+}$ single-ion anisotropy \cite{Wills:GSO,Glazkov:ESR} and
exchange interactions beyond nearest-neighbor \cite{Wills:GSO,Enjalran:GTO,Cepas:MFT}, 
one could in principle follow the well-trodden road of solid state physics and conventional magnetism: 
with the Hamiltonian and consequential ground state known, identify the
long wavelength excitations and, by second-quantizing them, calculate the low-temperature 
thermodynamic quantities.  It turns out that this program was carried out 
in a prior work \cite{DelMaestro:GTO} for a quantum version of a simple pyrochlore lattice model
with nearest-neighbor antiferromagnetic exchange plus long-range dipolar interactions \cite{Raju:GTO,PC}.
What was found in Ref.~[\onlinecite{DelMaestro:GTO}] is that 
{\it all} spin wave excitations of the Heisenberg pyrochlore antiferromagnet
 are pushed up in energy by the dipolar interactions and, as a result, all thermodynamic 
quantities show exponential temperature dependence, $\sim \exp(-\Delta/T)$, 
at low temperatures\cite{DelMaestro:GTO}.  The following question thus arises:
\begin{center}
\noindent
{\it
Do the $C_v\sim T^2$ results of Ref.~[\onlinecite{Bonville:GSO1}] for 
Gd$_2$Sn$_2$O$_7$ contradict the theoretical prediction of Ref.~[\onlinecite{DelMaestro:GTO}],
and are the magnetic excitations of Gd$_2$Sn$_2$O$_7$ truly unconventional?  
}
\end{center}
This is the question that we ask, and aim to answer in this paper.

\begin{figure}[t]
    \includegraphics[scale=0.40]{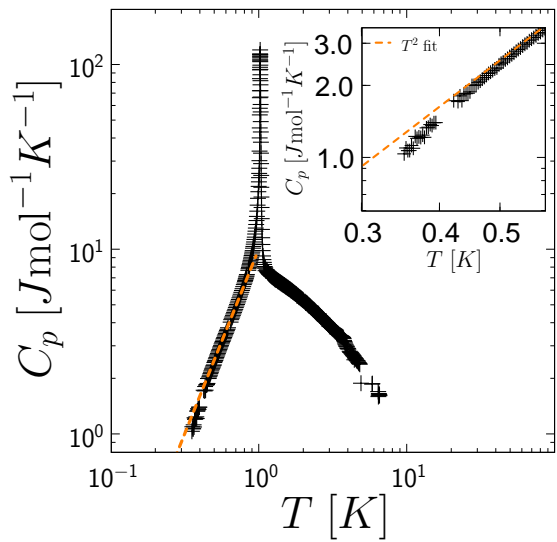}
    \caption{\label{fig:logCvlogT} (Color online) The specific heat of Gd$_2$Sn$_2$O$_7$ as a function of temperature,
	with an inset showing an enlargement of the low temperature region from Bonville 
	\textit{et al.}\cite{Bonville:GSO1} plotted on a logarithmic scale.  
	The dashed line shows the result of a relatively 
	successful $T^2$ fit below 0.75~K. [Data were generously provided by P. Bonville].}
\end{figure}

To spell out the question above more specifically, we show in Fig.~1 the specific heat data on 
Gd$_2$Sn$_2$O$_7$ reproduced from Ref.~[\onlinecite{Bonville:GSO1}].
The $T^2$ behavior (dashed line in main panel) 
ranges from $T_{\rm low}\sim 0.5~{\rm K}$ to $T_{\rm up} \sim 0.8 ~{\rm K}$. 
The $T_{\rm up}$ is very close to the critical temperature, and one does not expect on general 
grounds (non-critical) power-law behaviors reflecting excitations out of the ground state
to extend so close to the phase transition.
Secondly, the temperature $T_{\rm low}$, when compared with the 
results of Ref.~[\onlinecite{DelMaestro:GTO}], is {\it high} compared
to the temperature regime where we expect second-quantized (spin wave) excitations to describe
this system. Finally, and this is the key aspect of the data that prompted the present work,
we note that the $C_v$ data at $T\lesssim 0.5~{\rm K}$ progressively droop below the
dashed $C_v \sim T^2$ behavior. This is emphasized in the inset of Fig.~1.
Incidentally, we note from this plot that the $C_v\sim T^2$ behavior
does not provide a particularly good fit of the data for $T\in [350,800]$~mK.
The crux of the argument presented in this paper is that (i)
the $T^2$ power law between $T_{\rm low}$ and $T_{\rm up}$ is not a reflection of the
low-energy properties of Gd-based antiferromagnetic pyrochlores and, most importantly,
(ii) the behavior exhibited by $C_v$ below $T_{\rm low} \sim 0.5 ~{\rm K}$ (inset of Fig.~1) 
is a signature that
the system is progressively entering a low-temperature regime characterized by exponentially 
activated spin excitations over a gap originating from both 
magnetic dipole-dipole interactions and single-ion anisotropy.
 We show below via calculations that expand on the authors' previous work,
Ref.~[\onlinecite{DelMaestro:GTO}], that 
the specific heat data of Fig.~1 can be reasonably well described by such gapped magnetic excitations.
In other words, we assert that the bulk thermodynamic properties of 
Gd$_2$Sn$_2$O$_7$, revealed by data like that shown in Fig.~1, are compatible with a conventional
semi-classical long-range ordered phase.  We suggest that specific heat measurements below
$T_{\rm low}$ and down to 100~mK could be used to confirm or disprove our proposal.
The rest of the paper is organized as follows.
We first present a model for exchange and dipole coupled spins on the pyrochlore 
lattice in the presence of a crystal field inducing single ion spin anisotropy.
  The Hamiltonian is decoupled via
linear spin wave theory, and expressions for the quantum fluctuations and low temperature
thermodynamic properties are calculated.  We then
investigate the effects of second and third nearest neighbor magnetic exchange on the gap to spin
wave excitations.  Comparing the specific heat calculated in spin wave theory for Gd$_2$Sn$_2$O$_7$ 
to that measured in the experiments of Bonville \textit{et al.} \cite{Bonville:GSO1}, we use a
maximum likelihood estimator to determine a set of further neighbor couplings which may be 
present in the material. 
Finally, we identify the zero temperature quantum fluctuations 
and present a schematic phase diagram of Gd$_2$Sn$_2$O$_7$.

\section{Model Hamiltonian}
\label{sec:ModelHamiltonian}

As reviewed in the previous section,
the pyrochlore lattice has evinced much experimental and 
theoretical interest due to the large degree of
geometrical frustration arising from a structure consisting of
 corner sharing tetrahedra.  The cases of
Gd$_2$Ti$_2$O$_7$ and Gd$_2$Sn$_2$O$_7$ are 
somewhat special, in that the $\mathrm{S}=7/2$ Gd$^{3+}$ 
S-state ion should have a relatively small intrinsic anisotropy
when compared to other R$_2$Ti$_2$O$_7$ pyrochlore oxides.
It is known that the titanate (Gd$_2$Ti$_2$O$_7$)
has a complicated low temperature multi-$\vec{k}$ magnetic structure
\cite{Stewart:GTO}. 
However, recent neutron scattering \cite{Wills:GSO,Bonville:GSO1} 
and electron spin resonance \cite{Glazkov:ESR} experiments performed on 
gadolinium stanate (Gd$_2$Sn$_2$O$_7$) indicate that this material exhibits 
a $\vec{k}=0$ long range ordered state
below $\sim 1~{\rm K}$.  As such, Gd$_2$Sn$_2$O$_7$ should be reasonably
well described by a general two-body spin interaction Hamiltonian which includes 
predominant isotropic magnetic exchange interactions up to at 
least third nearest neighbor and anisotropy in the form of interactions with the local 
crystal field as well as long range dipole-dipole 
interactions~\cite{Raju:GTO,PC,Enjalran:GTO,Wills:GSO,Cepas:MFT,Cepas:MC}.

\begin{figure}[t]
    \includegraphics[scale=0.35]{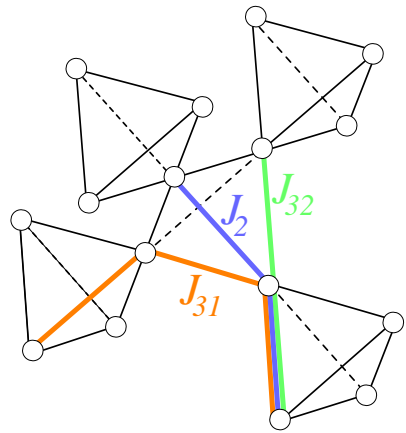}
    \caption{\label{fig:J3ExPaths} (Color online) A schematic portion of the pyrochlore lattice, detailing the paths
	which correspond to second nearest neighbor and two types of third nearest neighbor exchange interactions.}
\end{figure}

Such a Hamiltonian can be written as
\begin{equation}
\mathcal{H} = \mathcal{H}_{\rm ex} + \mathcal{H}_{\rm dd} + \mathcal{H}_{\rm cf},
\label{eq:H}
\end{equation}
where the exchange, dipole-dipole and crystal field terms are given by
\begin{subequations}
\begin{eqnarray}
\label{eq:Hex}
\mathcal{H}_{\rm ex} &=& -\frac{1}{2}\sum_{i,a}\sum_{j,b} J_{ab}(\vec{R}^{ij}_{ab}) 
			\,     \vec{S}_a(\vec{R}^i) \cdot \vec{S}_b(\vec{R}^j) \\
\label{eq:Hdd}
\mathcal{H}_{\rm dd} &=& \frac{D_{\rm dd}}{2}\sum_{i,a}\sum_{j,b} \left\{ \frac{\vec{S}_a(\vec{R}^i) 
                     \cdot \vec{S}_b(\vec{R}^j)}{|\vec{R}^{ij}_{ab}|^3} \right. - \nonumber \\ 
				 && \left. \quad\quad\quad 3 \frac{[\vec{S}_a(\vec{R}^i)\cdot \vec{R}^{ij}_{ab}]\; 
					       [\vec{S}_b(\vec{R}^j)\cdot \vec{R}^{ij}_{ab}]}{|\vec{R}^{ij}_{ab}|^5}
						   \right\} \\
\label{eq:Hcf}
\mathcal{H}_{\rm cf} &=& \sum_{i,a} \sum_{\ell,m}B^m_\ell\hat{O}^m_\ell[\vec{S}_a(\vec{R}^i)]
\end{eqnarray}
\end{subequations}
with the factors of $1/2$ having been included to avoid double counting.  
The various conventions used in
Eqs.~(\ref{eq:Hex}) to (\ref{eq:Hcf}) are as follows: $\vec{S}_a(\vec{R}^i)$ is the spin located on one of
$N$ tetrahedra identified by the face centered cubic (FCC) Bravais lattice vector $\vec{R}^i$ and the site by one of four 
tetrahedral sublattice vectors $\vec{r}_a$.  
$\vec{S}_a(\vec{R}^i)$ is assumed to be a full O(3) operator satisfying 
$\vec{S}_a(\vec{R}^i)\cdot\vec{S}_a(\vec{R}^i) = \mathrm{S}(\mathrm{S}+1)$. $J_{ab}(\vec{R}^{ij}_{ab})$ 
gives the value of the isotropic Heisenberg exchange interaction between two spins separated by
$\vec{R}^{ij}_{ab} = \vec{R}^j + \vec{r}_b - \vec{R}^i - \vec{r}_a$, with a negative sign
corresponding to antiferromagnetic interactions.  In this study, we focus on the 
Gd$_2$Sn$_2$O$_7$ material, 
and thus consider a fixed value of $J_1 = 3\Theta_{CW} / [z\mathrm{S}(\mathrm{S}+1)] = -0.273$~K where the 
Curie-Weiss temperature is $\Theta_{CW} = -8.6$~K and $z=6$ is the coordination number 
on the pyrochlore lattice \cite{Bonville:GSO1}. 
We treat the exchange interactions $J_2$ and $J_{31}$ beyond nearest neighbors
as parameters to be adjusted below to produce agreement with experimental 
(specific heat) measurements on Gd$_2$Sn$_2$O$_7$. Following the 
approach of Wills \textit{et al.}\cite{Wills:GSO} 
we treat the two possible third nearest neighbor (NN) exchange paths 
$J_{31}$ and $J_{32}$, known to be present in the pyrochlores\cite{Wills:GSO,Kenedy:TinPyro}, 
separately  (see Fig.~\ref{fig:J3ExPaths}). 
With the expectation that $J_{32} \ll J_{31}$, we henceforth set $J_{32} = 0$.
The strength of the dipole interaction is given by  $D_{\rm dd} = \mu_0 (g\mu_B)^2 / 4\pi$.
At nearest neighbor distance,
$R_{\rm nn} = a\sqrt{2}/4 = 3.695$~\AA~, where $a = 10.45$~\AA~ is the size of the cubic unit cell, 
$D_{\rm dd}/{R_{nn}}^3$ is approximately 15\% of the 
exchange energy $J_1$.  The crystal field Hamiltonian is written as an 
expansion of Stevens operators, $\hat{O}^m_\ell$, that transform like the real
tesseral harmonics \cite{Hutchings:CF}.  The number of terms in the expansion is 
strongly constrained by symmetry and, from recent 
electron spin resonance (ESR)
measurements \cite{Glazkov:ESR}, the values of $B_2^0$ and $B_4^0$ have been 
estimated at $(47 \pm 1)$~mK and $(0.05 \pm 0.02)$~mK, respectively.  
Here, we only consider the 
dominant lowest order term in the expansion of $\mathcal{H}_{\rm cf}$, 
$B_2^0$, which contributes energetically on equal footing with
the dipole interactions, and leave the inclusion of higher order corrections to a future study.  
Writing the Steven's operators in terms of angular momentum operators
\cite{Jensen:REM},
the crystal field part of the Hamiltonian ${\cal H}$, ${\cal H}_{\rm cf}$, is:
\begin{equation}
\mathcal{H}_{\rm cf} = -4N B_2^0\mathrm{S}(\mathrm{S}+1) + 3 B_2^0\sum_{i,a} 
				    \left[\vec{S}_a(R^i) \cdot \hat{z}_a \right]^2,
\label{eq:HcfSpin}
\end{equation}
where the four unit vectors $\hat{z}_a$ describe the local
 $\langle 111 \rangle$ direction for each site on a tetrahedron.  The
conventions and definitions used in this study for all vectors and lengths on the pyrochlore lattice are 
given in Table~1 of a previous work by one of the authors \cite{Enjalran:TTO}.

\begin{figure}[t]
    \includegraphics[width=3.3in]{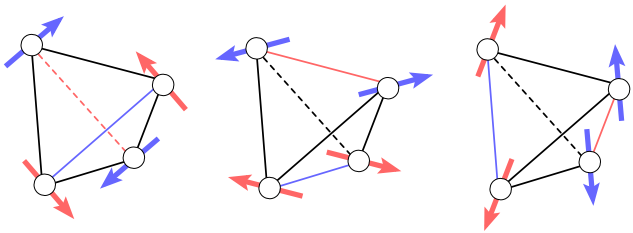}
  \caption{\label{fig:PC} 
(Color online) Six degenerate Palmer-Chalker (PC)
ground states (reverse spins for other three) for spins on a
single tetrahedron. The ground states are characterized by
each spin being parallel to an edge of the tetrahedron that
does not intersect its vertex and all are tangent to the sphere
which circumscribes it. The net magnetic moment on each
tetrahedron is identically zero.}
\end{figure}

We are interested in the effects of the low energy excitations (spin waves) on the thermodynamic properties
of a real material described by Eq.~(\ref{eq:H}).  
At zero temperature, we assume that the system is in one of 
the six discrete Palmer-Chalker (PC) $\vec{k} = 0$ ground states \cite{PC} 
depicted in Fig.~\ref{fig:PC}.
We have confirmed by direct numerical simulations that the classical
zero temperature ground state in a model with
nearest-nearest neighbor antiferromagnetic exchange and
long-range dipolar interactions at the level of 10 -- 20\% of the exchange is
the PC ground state. See also Refs.~[\onlinecite{Enjalran:GTO,Cepas:MFT,Cepas:MC}].
The use of the PC state for Gd$_2$Sn$_2$O$_7$ 
is supported by recent powder neutron scattering experiments \cite{Wills:GSO} 
where the magnetic diffraction pattern was compared to the expected result from multiple candidate 
ground states.  Our approach below will be to analyze the stability of this ground state and its accompanying 
excitations by investigating the role of quantum fluctuations in reducing the fully polarized classical 
spin value of $\vec{S}_{\rm cl} = (0,0,\mathrm{S})$.  This can be accomplished by changing the axis of quantization 
from the global $z$-direction (an arbitrary choice) to a local axis described by a triad of unit vectors 
$\{\hat{n}^u_a\}$.  This triad is defined such that the locally quantized spin, denoted by a tilde,
 is related to the spin operator in the Cartesian lab frame via a rotation
\begin{eqnarray}
\vec{S}_a(\vec{R}^i) &=& \mathsf{R}(\theta_a,\phi_a)\tilde{\vec{S}}_a(\vec{R}^i) \nonumber \\
				     &=& \sum_{u} \tilde{S}^u_a(\vec{R}^i)\hat{n}^u_a
\label{eq:Slocframe}
\end{eqnarray}
where $u$ runs over the Cartesian indices $\{x,y,z\}$ and $\hat{n}^z_a$ points in the direction of each 
classical spin. Using Eq.~(\ref{eq:Slocframe}) we can rewrite Eq.~(\ref{eq:H}) in a much more compact form
\begin{equation}
\mathcal{H} = -\frac{1}{2}\sum_{a,b}\sum_{i,j}\sum_{u,v}
\tilde{S}^u_a(\vec{R}^i) \mathcal{J}^{uv}_{ab}(\vec{R}^{ij}_{ab}) \tilde{S}^v_b(\vec{R}^j),
\label{eq:Hlf}
\end{equation}
where we have neglected a constant term and 
\begin{eqnarray}
\mathcal{J}^{uv}_{ab}(\vec{R}^{ij}_{ab}) &=& J_{ab}(\vec{R}^{ij}_{ab})\hat{n}^u_a\cdot\hat{n}^v_b \nonumber \\
&&-\ \frac{D_{dd}}{\left|\vec{R}^{ij}_{ab}\right|^3} \left[ \hat{n}^u_a \cdot \hat{n}^v_b
-  3(\hat{n}^u_a \cdot \hat{R}^{ij}_{ab})(\hat{n}^v_b\cdot\hat{R}^{ij}_{ab}) \right] \nonumber \\
&&-\ 6B_2^0 (\hat{n}^u_a \cdot \hat{z}_a)(\hat{n}^v_b \cdot \hat{z}_b)
\delta_{a,b}\delta_{u,v}.
\label{eq:fullJ}
\end{eqnarray}
Having manipulated our Hamiltonian into a more manageable form,  we employ 
in the next section the methods of linear spin wave theory to diagonalize Eq.~(\ref{eq:Hlf}).

\section{Linear spin wave theory}
\label{sec:LSWT}

In a previous study\cite{DelMaestro:GTO}, which we henceforth refer to 
as DG, we presented the diagonalization of 
$\mathcal{H}_{\rm ex} + \mathcal{H}_{\rm dd}$ on the pyrochlore lattice via a 
Holstein-Primakoff\cite{HolstPrim} spin wave expansion to order $1/\mathrm{S}$, through the introduction of
bosonic spin deviation (magnon) creation (annihilation) operators $c^\dag_a$ ($c^{\phantom\dag}_a$).  
The Ewald
summation technique \cite{ewald} was used to calculate the Fourier transform of the infinite 
range dipole-dipole
interaction matrix.  The calculations of DG can be straightforwardly
generalized to include the effects of the crystal field by shifting the diagonal 
spin interaction matrix elements ($\mathsf{A}_{\alpha\alpha}(\vec{k})$ and 
$\mathsf{B}_{\alpha\alpha}(\vec{k})$ of Eq.~(16) in DG) by a term proportional to $B_2^0$ 
(see Eq.~(\ref{eq:fullJ})). 
The result, after the usual Bogoliubov diagonalization procedure,
 is a Bose gas of non-interacting spin waves described by
\begin{equation}
\mathcal{H} = \mathcal{H}^{(0)} + \sum_{\vec{k}}\sum_a \varepsilon(\vec{k})\left[
a^\dag_a(\vec{k})a^{\phantom\dag}_a(\vec{k}) + \frac{1}{2}\right],
\label{eq:SWHam}
\end{equation}
where the summation is over all wavevectors in the first Brillouin zone 
(BZ) of the FCC lattice.  The dispersion relations for the spin wave modes, 
$\varepsilon_a(\vec{k})$, are calculated from the spectrum of the Bogoliubov transformation.  
Physically, they are 
identical to $\varepsilon_a(\vec{k})=\hbar \omega_a(\vec{k})$ where
$\omega_a(\vec{k})$ are the classical excitation frequencies obtained by linearizing the classical 
equations of motion for interacting magnetic dipoles or rotors \cite{DelMaestro:MSC}.

In a real magnet, spin wave fluctuations with dispersion $\varepsilon_a(\vec{k})$
 raise the classical ground state energy and reduce the staggered magnetic
moment per spin from its classical value of $\mathrm{S}$. 
From  Eq.~(\ref{eq:SWHam}),
the contribution to the ground state energy is given by
\begin{equation}
    \Delta \mathcal{H}^{(0)} = \frac{1}{2}\sum_{\vec{k}}\sum_{a} \varepsilon_a (\vec{k}).
    \label{eq:dH}
\end{equation}
The full spectrum of the Holstein-Primakoff transformation can be used to calculate
the reduction in the staggered magnetization
\begin{equation}
    \Delta S = \frac{1}{2}\left(\frac{1}{8\mathrm{N}} 
    \sum_{\vec{k}}\mathrm{Tr}[\mathsf{Q}^\dag\mathsf{Q}] - 1 \right),
    \label{eq:dS}
\end{equation}
where $\mathsf{Q}$ is the $8 \times 8$ hyperbolically normalized matrix of eigenvectors such
that $\mathrm{Tr}\; (\mathsf{Q}^\dag\mathsf{H}\mathsf{Q}) = \sum_a \varepsilon_a
(\vec{k})$ and $\mathsf{H}$ is the $8 \times 8$ block matrix Hamiltonian 
[see DG Eqs.~(16) and (19)].
$N$ is the number of tetrahedra on a pyrochlore lattice with periodic boundary conditions.

At low temperatures ($k_B T < \varepsilon_a(\vec{k})$) expressions for the specific heat at
constant volume, $C_v$, and staggered magnetization,
$m=S-\Delta S$, can be derived from the classical
partition function $\mathcal{Z} = \mathrm{Tr}\; [\exp(-\beta \mathcal{H})]$ corresponding to 
Eq.~(\ref{eq:SWHam}).  Using Eqs.~(\ref{eq:SWHam}) to (\ref{eq:dS}), we find (see DG)
\begin{eqnarray}
\label{eq:Cv}
C_v &=& \beta^2 \sum_{\vec{k}}\sum_{a}[\varepsilon_a(\vec{k})n_B(\varepsilon_a(\vec{k}))]^2 
\exp[\beta\varepsilon_a(\vec{k})], \\
\label{eq:M}
m &=& \mathrm{S} + \frac{1}{2} \nonumber \\
&& -\frac{1}{8\mathrm{N}}\sum_{\vec{k}}\sum_{a}[\mathsf{Q}^\dag\mathsf{Q}]_{aa}
	[1+n_B(\varepsilon_a(\vec{k}))],
\end{eqnarray}
where $\beta$ is the inverse temperature and $n_B(\varepsilon_a(\vec{k})) =
1/(\mathrm{e}^{\beta\varepsilon_a(\vec{k})}-1)$ is the Bose distribution function.

\section{Results}
Having determined the expressions for the zero temperature quantum fluctuations, as well as 
finite temperature thermodynamic relations of a dipolar Heisenberg model with crystal field 
interactions on the pyrochlore lattice, we may now study the quantitative effects of perturbative 
$J_2$ and  $J_{31}$ for fixed $J_1$ and $B_2^0$.   The limit of stability of the proposed 
PC classical ground states on the pyrochlore lattice can be investigated by searching for 
soft modes ($\varepsilon_a(\vec{k}) \to 0$) at some wavevector $\vec{k}$.

\subsection{Second and third NN exchange}
\begin{figure}[t]
\includegraphics[scale=0.40]{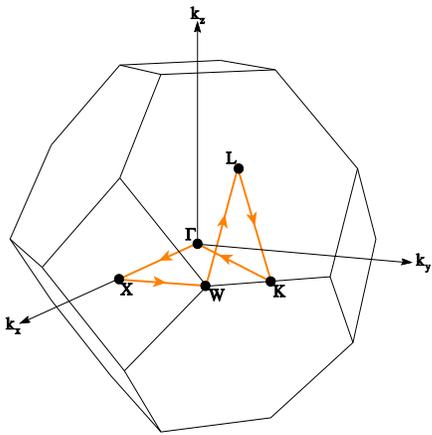}
\caption{(Color online) The first Brillouin zone of the pyrochlore lattice, showing the five high
symmetry points $\Gamma = (0,0,0)$, $X = 2\pi/a (1,0,0)$, $W=2\pi/a(1,1/2,0)$,
$L=2\pi/a(1/2,1/2,1/2)$ and $K = 2\pi/a(3/4,3/4,0)$ and the path in $k$-space along which spectra
are plotted in this study. \label{fig:FCCBZ}}
\end{figure}

If we ignore all interactions except nearest neighbor isotropic antiferromagnetic exchange, it has been known for
some time\cite{Villain} that both classical \cite{Moessner:Pyro} and quantum \cite{Canals:Pyro} 
spins on the pyrochlore
lattice fail to develop conventional long range magnetic order at nonzero temperature.
There exists two unbroken symmetries in the ground state manifold \cite{Reimers-MFT,Moessner-CJP}
corresponding to lattice translation and time reversal.
The excitation spectrum consists of two sets of two 
degenerate modes, the first being soft over the entire Brillouin zone, (see Fig.~\ref{fig:FCCBZ}) and 
the second are acoustic and linear along $\Gamma \to X$ leading to divergent quantum fluctuations.  
The inclusion of both dipole-dipole, and crystal field interactions break the rotational symmetry and lift 
\textit{all} excitations to finite frequency \cite{DelMaestro:GTO}. 

\begin{figure}[t]
\includegraphics*[scale=0.40]{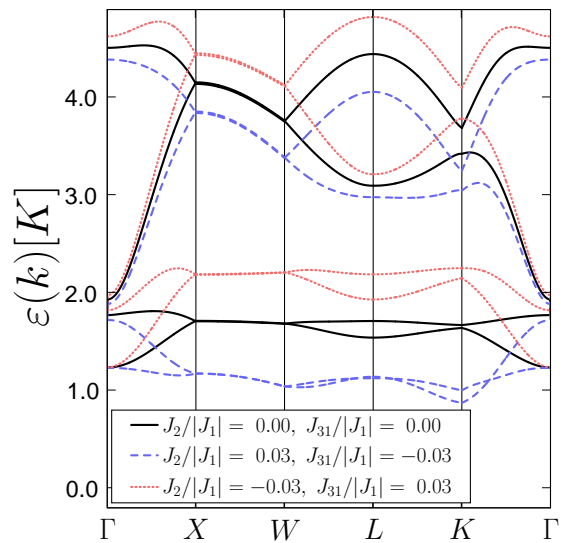}
\caption{(Color online) The spin wave excitation spectrum in kelvin for $(J_2/|J_1|,J_{31}/|J_1|)$ equal to $(0.0,
0.0)$, $( 0.03,-0.03)$ and $(-0.03, 0.03)$ plotted along a high-symmetry path in
the first Brillouin zone of the FCC lattice.  For the three parameter sets
shown there exists a finite gap to spin wave excitations throughout the zone.
\label{fig:hspSpectra}}
\end{figure}

Fig.~\ref{fig:hspSpectra} shows the dispersion spectrum of the four spin wave modes for three values of 
$J_2$ and $J_{31}$ along the high symmetry path described in Fig.~\ref{fig:FCCBZ}. 
We do indeed observe only optical modes, with degeneracy preserved along $X \to W$ for $J_2 = J_{31} = 0$.  
The qualitative behavior of $\varepsilon_a(\vec{k})$ with varying $J_2$ and $J_{31}$ 
also seems to confirm the naive expectation that even perturbatively small third nearest 
antiferromagnetic neighbor exchange should 
reduce the stability of the PC states~\cite{Enjalran:GTO,Cepas:MFT,Wills:GSO}. 
 It also appears that the minimum excitation gap does not always 
occur at $\vec{k}=0$ indicating the underlying presence of a finite wavevector instability
as $J_2$ and $J_{31}$ are tuned away from $J_2=J_{31}=0$.

\begin{figure}[t]
\includegraphics[scale=0.40]{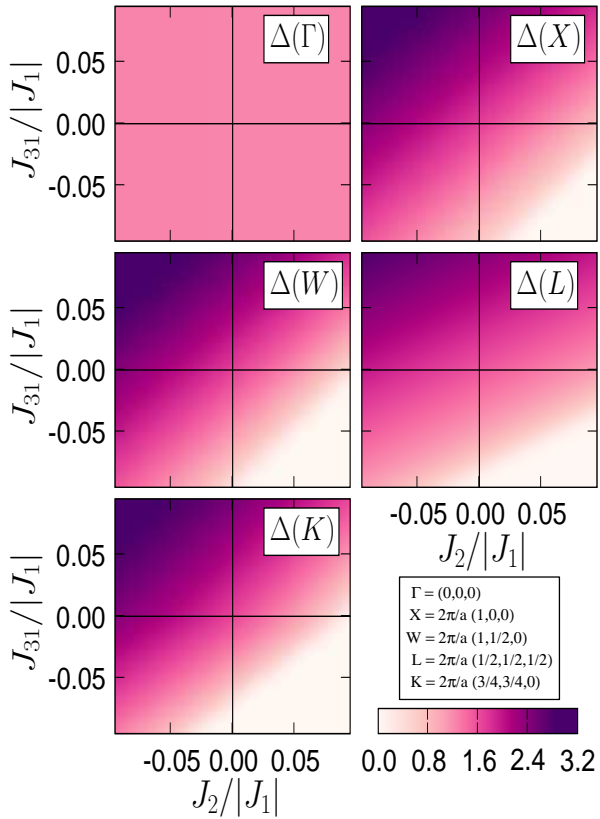}
\caption{(Color online) The magnitude of the lowest spin wave excitation
energy $\Delta(k)$ in kelvin at various high symmetry points in the first BZ plotted 
in the $J_2 - J_{31}$ plane.  All panels are plotted with the same color scale making 
the slight variations in the gap at $\Gamma = (0,0,0)$ difficult to discern.  
A soft mode instability occurs only for antiferromagnetic third NN coupling $J_{31}$
\label{fig:hspKPoints}}
\end{figure}

\begin{figure}[t]
\includegraphics*[scale=0.4]{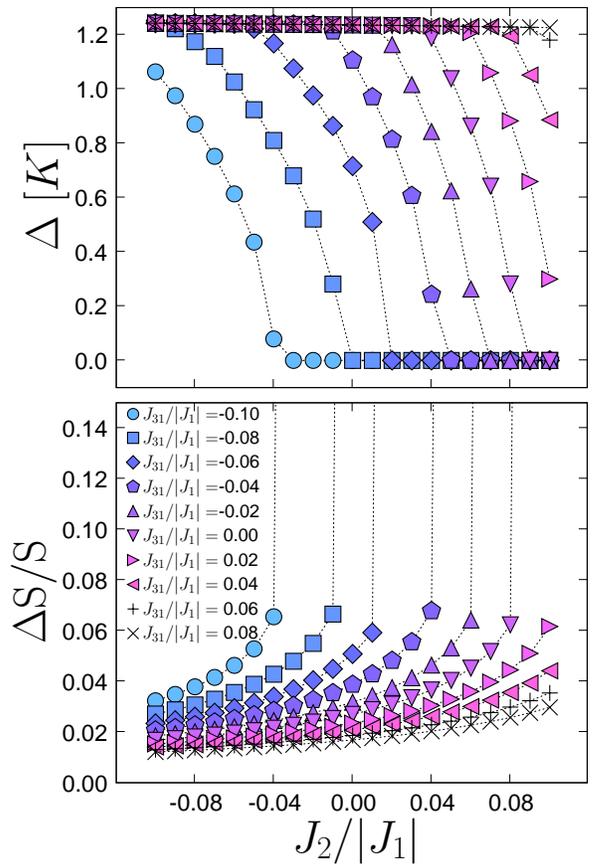}
\caption{(Color online) 
The spin wave excitation gap in kelvin ($\Delta$, top panel) 
found by determining the minimum energy over $19^3$ discrete points in the first BZ, 
and the reduction in sublattice magnetization ($\Delta\mathrm{S} / \mathrm{S}$, bottom panel) 
for various values of $J_{31}/|J_1|$ plotted against $J_2/|J_1|$. 
The jump in $\Delta \mathrm{S}/\mathrm{S}$ occurs once the limit of stability of the
$\vec{k}=0$ Palmer-Chalker  ground state is reached.
Both panels share a common legend.  
\label{fig:gapMagVer}}
\end{figure}

The spin wave energy gap can be analyzed more quantitatively by defining
\begin{subequations}
\begin{eqnarray}
\Delta (\vec{k}) &\equiv& \min_{a} \left[\varepsilon_a(\vec{k})\right] \\
\Delta &\equiv& \min_{\vec{k}} \left[ \Delta(\vec{k})\right].
\label{eq:gap}
\end{eqnarray}
\end{subequations}
The value of $\Delta(\vec{k})$ can be investigated as a function of $J_2$ and $J_{31}$ at each of the high
symmetry points described above.  As we vary $J_2$ and $J_{31}$ through some critical values,
instabilities first appear at these wavevectors of high symmetry.  The resulting
gap values are shown in Fig.~\ref{fig:hspKPoints}.
Although $\Delta(\Gamma) > 0$ for all values of $J_2$ and $J_{31}$ studied here, the region of stability of the
PC states is defined by the observed appearance of soft modes, 
at $\vec{k} = K = 2\pi/a(3/4,3/4,0)$ for ferromagnetic 
(positive)
$J_2$ and antiferromagnetic 
(negative)
$J_{31}$.
Performing a search for the minimum value of the gap
over the entire Brillouin zone ($\Delta$) at each value of 
$J_2$ and $J_{31}$ confirms that the instability first appears at the $K$-point.  
The effect of perturbative second and third NN exchange interactions on the global minimum energy 
gap (Eq.~(\ref{eq:gap})) along with the corresponding magnitude of spin fluctuations 
$\Delta \mathrm{S} / \mathrm{S}$ (Eq.~(\ref{eq:dS})) is shown in Fig.~\ref{fig:gapMagVer}. 
Here we observe that upon reaching a pair of critical values for $J_2/|J_1|$ and $J_{31}/|J_1|$, the
excitation gap is suppressed to zero (top panel), and divergent spin fluctuations ensue (bottom panel).
The values of $J_2$ and $J_{31}$ corresponding to $\Delta \to 0$ 
can be identified, and are best described by the linear relationship
$J_{31} = 0.750J_2 - 0.077|J_1|$.  This line defines the phase boundary between a sector of 
stability for the $\vec{k}=0$ PC ground
states, and a region characterized by instabilities at finite wavevector.  In addition, this line 
corresponds to the white regions in Fig.~\ref{fig:hspKPoints} where $\Delta \to 0$, 
and thus defines the limit of applicability of the spin wave calculation around
the PC ground state described in Section \ref{sec:LSWT}. 

\begin{figure}[t]
\includegraphics[scale=0.40]{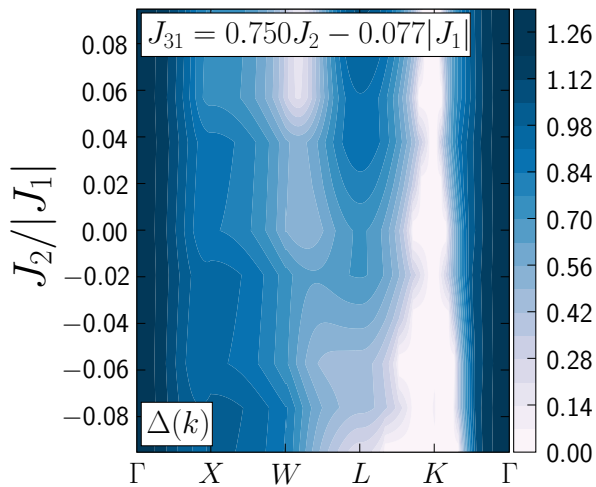}
\caption{(Color online) The lowest excitation energy $\Delta(k)$ in kelvin plotted along the high symmetry
path discussed in the text with $(J_2,J_{31})$ fixed to reside on the phase 
boundary parameterized by $J_{31} = 0.750 J_2 - 0.077$. 
The location of the zero modes in $k$-space indicate that the instability out of the PC states occurs
via excitations described by ${\vec k}=2\pi/a(3/4,3/4,0)$.
\label{fig:J2hspSpectra}}
\end{figure}

Plotting $\Delta(\vec{k})$ along $\Gamma \to X \to W \to L \to K \to \Gamma$
with $J_{31}$ pinned to this phase boundary leads to the spectrum shown in Fig.~\ref{fig:J2hspSpectra}.
It is apparent from this result, that once the value of the third NN exchange constant has been set at a
suitably antiferromagnetic value, altering the second NN exchange constant, has a relatively limited effect 
on the gap and on the consequential proliferation of quantum fluctuations about the classical ground state.

To summarize, the effects of perturbative second and third NN exchange interactions on the 
appearance of soft modes and their accompanying quantum fluctuations in a model of a dipolar coupled
antiferromagnetic Heisenberg pyrochlore with single-ion anisotropy 
is globally illustrated in Figs.~\ref{fig:hspSpectra}-\ref{fig:J2hspSpectra}.  
Such a model should well characterize the low temperature behavior of Gd$_2$Sn$_2$O$_7$, and we 
next apply these tools with the goal of searching for the unconventional
spin excitations believed to be present in this material
on the basis of the unconventional $C_v\propto T^2$ specific heat in the temperature
range $[350,800]$ mK.

\subsection{The case of Gd$_2$Sn$_2$O$_7$}

As described in the Introduction, recent studies\cite{Bonville:GSO1,Bonville:GSO2} of
Gd$_2$Sn$_2$O$_7$ have reported, on the basis of $\mu$SR measurements,
evidence for Gd$^{3+}$ spin dynamics well below 0.9~K, as well as  
suggesting that the low temperature specific heat is accurately 
described by an anomalous $T^2$ power
law.  This is in stark contrast with
the expected $T^3$ behavior for a three dimensional
antiferromagnet, with possible exponential suppression 
at a temperature below a characteristic excitation gap.  

\begin{figure}[t]
\includegraphics[scale=0.40]{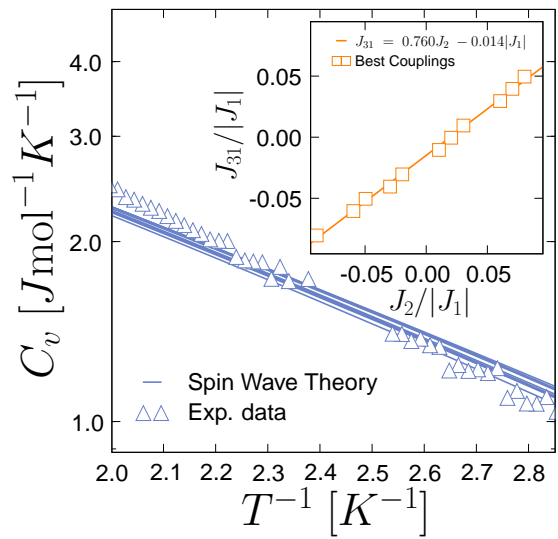}
\caption{(Color online) The low temperature specific heat from 
Ref.~[\onlinecite{Bonville:GSO1}] plotted against inverse temperature on a logarithmic
scale. The lines are the calculated value of the specific heat using the
ten best fit values for $(J_2/|J_1|, J_{31}/|J_1|)$ (as seen in the inset).
\label{fig:logCvvs1oT}}
\end{figure}

On the other hand, the long-range ordered state found by
neutron scattering in Gd$_2$Sn$_2$O$_7$ is that predicted by the simple model
of Eq.~\ref{eq:H} in Section III and discussed in 
Refs.~[\onlinecite{Raju:GTO,Cepas:MFT,Cepas:MC,Enjalran:GTO,PC}] and, consequently,
the low-temperature behavior of this material should be well described by linear spin wave theory.
Thus,  in an attempt to resolve the paradox offered by the $C_v\sim T^2$ behavior,
we have calculated the low temperature specific heat via Eq~(\ref{eq:Cv}) 
within the $J_2-J_{31}$ plane, and have performed a search
for the parameters which best reproduce the reported low temperature specific heat\cite{Bonville:GSO1}.  This
was accomplished by performing least squares linear fits of $\log C_v$ vs $1/T$ for $T < 0.5$~K 
between the experimental data and the spin wave specific heat for approximately 500 values of $J_2$ and $J_{31}$.  
A characteristic subset of the large number of performed fits are displayed in Fig.~\ref{fig:logCvvs1oT}.
\begin{figure}[t]
\includegraphics[scale=0.40]{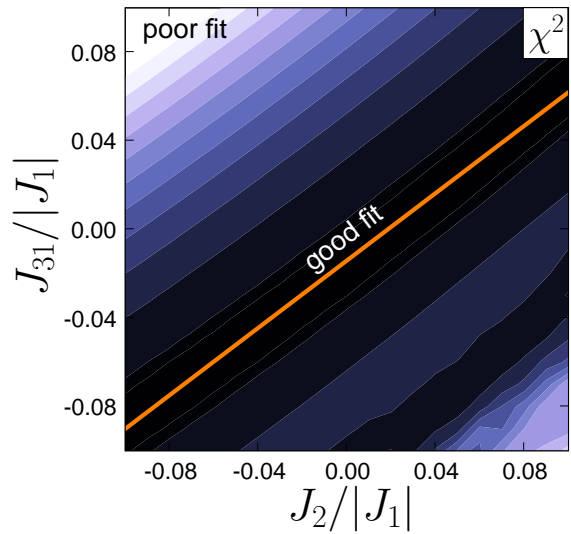}
\caption{(Color online) The value of the maximum likelihood estimator
($\chi^2$) in the $J_2 - J_{31}$ plane for a fit of the 
logarithm of the specific heat at low temperatures ($T < 0.5$~K) calculated
using linear spin wave theory and compared with the experimental data of
Ref.~[\onlinecite{Bonville:GSO1}].  
The parameters $(J_2/|J_1|,J_{31}/|J_1|)$ with the smallest $\chi^2$, giving the most effective fit, 
fall along the indicated line $J_{31} = 0.760 J_2 - 0.014|J_1|$.
\label{fig:chi2J2J31}}
\end{figure}
The values of $J_2$ and
$J_{31}$ which provided the best fit to the experimental data can be quantified by defining a 
maximum likelihood estimator $\chi^2$ which is shown in Fig.~\ref{fig:chi2J2J31}.
It is important to note that the fits of the specific heat, $C_v$, discussed here, 
were done with an {\it absolute} dimensionfull
scale, and thus no vertical adjustment of the experimental data was allowed.
The comparisons allow only for adjustments of $J_2$ and $J_{31}$ which are
therefore {\it fine-tuning} effects. As such, it appears that a model which possesses solely 
nearest-neighbor exchange, long-range dipolar interactions and single-ion anisotropy 
{\it already} leads to a reasonable semi-quantitative description of the $C_v$ data below 500~mK.
This indicates that the temperature $T\sim 500$~mK corresponds to the upper temperature below 
which magnetic excitations become thermally activated.

The minimum of $\chi^2$ in the $J_2-J_{31}$ plane falls along the straight 
line $J_{31} = 0.760 J_2 - 0.014|J_1|$. 
This line of best fit also falls in a region of large stability (highly gapped spin wave excitations) 
for the classical PC ground states with respect to quantum fluctuations (see Fig.~\ref{fig:hspKPoints}).

The poorness of fit for simultaneously strong ferromagnetic second NN and antiferromagnetic third 
NN interactions or vice versa (top left, or lower right of Fig.~\ref{fig:chi2J2J31}), 
seems to indicate that is quite unlikely that 
 Gd$_2$Sn$_2$O$_7$ resides in these portions of the phase diagram. 
The parameters $J_2 = 0.02$ and $J_{31} = 0.0$ provide 
the \textit{best} empirical fit to the experimental specific heat data, 
although qualitatively similar fits are seen for all parameters which satisfy 
$J_{31} = 0.760 J_2 - 0.014|J_1|$.  
\begin{figure}[t]
\includegraphics[scale=0.40]{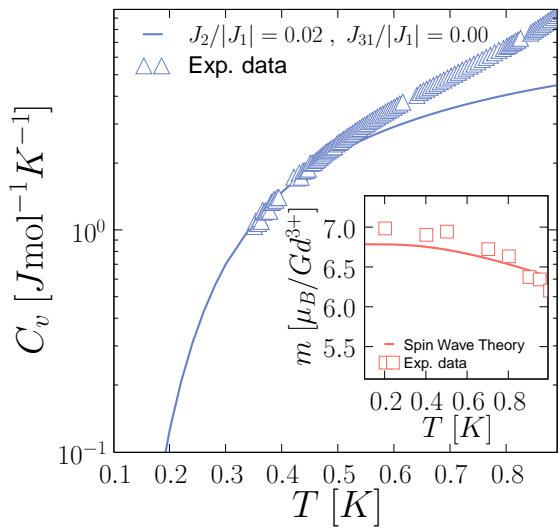}
\caption{(Color online) The low temperature specific heat 
(data from Ref.~[\onlinecite{Bonville:GSO1}]) versus temperature on a logarithmic
scale.  The calculated specific heat is plotted for the parameter set 
$J_2 = 0.02$, $J_{31} = 0.0$ which had the smallest $\chi^2$, 
although any values close to the line mentioned in the text give a
qualitatively very similar result. The agreement is quite good at low temperatures, where spin wave
theory should be applicable. The low temperature suppression of $C_v$ is
characteristic of gapped excitations, in contrast with previously reported power 
law behavior\cite{Bonville:GSO1,Bonville:GSO2}. The inset displays the temperature dependence of the 
Gd$^{3+}$ moment calculated using a Bose gas of excitations along with the value measured from 
$^{155}$Gd M\"{o}ssbauer measurements \cite{Bonville:GSO1}.
\label{fig:CvvsT}}
\end{figure}
Setting the parameters to these particular values, we display the spin wave specific heat, 
as well as the temperature dependence of the order parameter $m$ in Fig.~\ref{fig:CvvsT}.
Again, as shown in Fig.~\ref{fig:logCvvs1oT}, it is clear that at low temperatures, 
($T\le 370$~mK), the experimental specific heat data 
systematically falls below the gapped spin wave results.  This behavior is tentatively
consistent with the fact that spin wave theory produces a smaller value for the magnetization $m$ than
what is measured from M\"{o}ssbauer experiments.  However, we note that due to the 
intrinsic short dynamical time scale probed by
M\"{o}ssbauer measurements, the experimental data in the inset of Fig.~\ref{fig:CvvsT} may not reflect
the true value of the infinite-time order parameter.

It is perhaps worthwhile to make a few comments on the physical meaning
of the above fits.  Firstly, we note that
because of the weakly dispersive nature of the two lowest lying gapped magnon excitations
(see Fig.~\ref{fig:hspSpectra}), the temperature dependence of thermodynamics quantities
in the pyrochlore Heisenberg antiferromagnet plus dipolar interactions do not display 
the typical $C_v\sim T^3$ behavior as the temperature reaches approximately 0.5 K and exits
its characteristic low-temperature exponential behavior. 
In fact, such an observation was already
made in Ref.~[\onlinecite{DelMaestro:GTO}]
independently of any attempt to describe $C_v$ for Gd$_2$Sn$_2$O$_7$.
Secondly, the calculations presented here constitute a standard procedure for a system
with conventional long range magnetic order.
In this context, it is therefore
interesting to note that, contrary to the reported $T^2$ behavior, 
the experimental specific heat data are not only  
relatively well fit using the exponential spin wave form of Eq.~(\ref{eq:Cv}), 
but it appears to fall off even faster
than the exponentials considered at low temperatures.  
This would lend credence to the view that 
analyzing experimental data on a log-log scale over a limited
range, can lead to specious power law fits.
Hence, and on the basis of specific heat measurements 
\emph{alone} (i.e. without consideration of the $1/T_1$ $\mu$SR 
spin-lattice relaxation rate),
it would therefore appear that the suggestion of unconventional excitations in 
Gd$_2$Sn$_2$O$_7$ should be challenged by the principle of ``Ockham's razor''.
We are therefore led to suggest that the description of the specific heat $C_v$ in 
terms of an anomalous power law, $C_v\sim T^2$, in a {\it reduced} and {\it intermediate}
temperature range $T\in [350,800]$~mK does not provide a convincing indicator
for anomalous excitations out of the ground state of Gd$_2$Sn$_2$O$_7$.
Unlike the suggestion made in Ref.~[\onlinecite{Bonville:GSO2}] on the basis
of the temperature independence of the $1/T_1$ muon spin relaxation rate below
$T_c\sim 1~{\rm K}$, we have found a fully gapped
spin wave spectrum with no density of states at zero
energy. Hence, at this time, the microscopic origin of the
temperature independence of $1/T_1$ found below $T_c$ in
Gd$_2$Sn$_2$O$_7$ remains to be understood.

\subsection{Ground state properties}

\begin{figure}[t]
\includegraphics[scale=0.30]{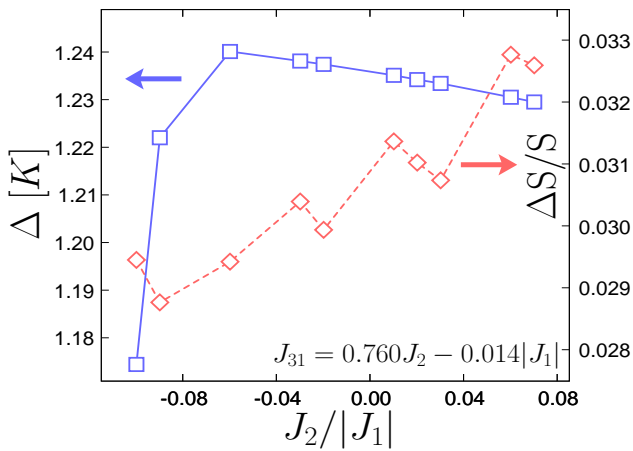}
\caption{(Color online) The global spin wave energy gap (left scale) and the
reduction in the staggered moment (right scale) along the line in the $J_2 -
J_{31}$ plane most closely corresponding to the parameter set of Gd$_2$Sn$_2$O$_7$. 
\label{fig:bestFitGapMag}}
\end{figure}

\begin{figure}[t]
\includegraphics[scale=0.40]{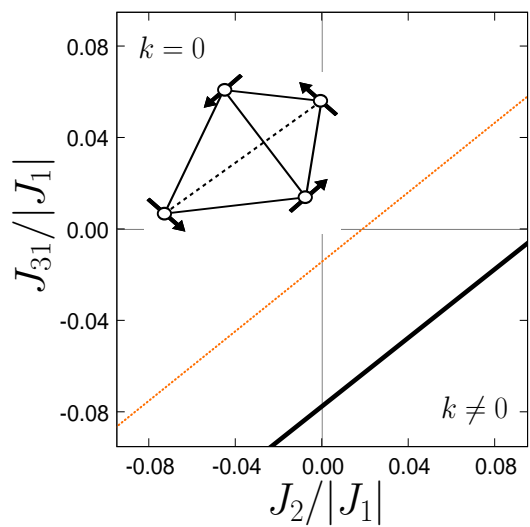}
\caption{(Color online) The $J_2 - J_{31}$ ground state phase diagram showing 
regions with order characterized by zero and non-zero wavevectors.  
The solid black line delineates the limit of stability of the Palmer-Chalker ground state.
The inset tetrahedron is tiled with one of the PC states, and the dashed line 
$J_{31} = 0.760 J_2 - 0.014|J_1|$
corresponds to the values of $J_2$ and $J_{31}$ which produced the best fits to experimental data.
\label{fig:J2J31PhaseDiagram}}
\end{figure}

In the previous section, we have identified a
relationship between the second $J_2$ and third $J_{31}$
NN exchange constants which best reproduce the
low temperature thermodynamic behavior in Gd$_2$Sn$_2$O$_7$.
  We now investigate the role of quantum fluctuations.
Fig.~\ref{fig:bestFitGapMag} displays both the minimum spin wave energy gap $\Delta$ and the reduction in the 
staggered moment $\Delta \mathrm{S} / \mathrm{S}$ along the line of best fit $J_{31} = 0.760 J_2 - 0.014|J_1|$.  
This result details the complicated relationship between the  value of the gap, and the stability of the 
ground state, i.e.  a decrease in the global spin wave energy gap (which may only occur at a
single $\vec{k}$-point) does not immediately trigger 
an increase of moderate quantum fluctuations.
Indeed, the opposite behavior is seen in Fig.~\ref{fig:bestFitGapMag}.  
In an exchange coupled
Heisenberg antiferromagnet on a non-Bravais lattice, the specifics 
of all relative energy scales come into play, and
one must not neglect the effects of weakly dispersing optical modes.
All the results presented here can be compiled into an effective 
schematic phase diagram for Gd$_2$Sn$_2$O$_7$.
Fig.~\ref{fig:J2J31PhaseDiagram} depicts the separatrix in the $J_2-J_{31}$ plane 
(solid line) that delineates the limit of stability of the $k=0$ PC state
against a soft mode characterized by wavevector $2\pi/a(3/4,3/4,0)$, i.e. the $K-$point.  
Possibly relevant values of $J_2$ and $J_{31}$ 
for which $\chi^2$ reaches it minimum value are shown as the  
parametric dashed line $J_{31}=0.760J_2-0.014 |J_1|$. 
We note that Fig.~\ref{fig:J2J31PhaseDiagram} shows a zero temperature phase diagram that
delineates the limit of stability of the $\vec{k}=0$ Palmer-Chalker ground state against
an instability at $2\pi/a(3/4,3/4,0)$.
A similar phase
diagram presented in Ref.~[\onlinecite{Wills:GSO}] (with the signs corresponding to FM and 
AF interactions reversed) gives the ordering wave vector of the
long range magnetic ordered state that first develops as the system is cooled down from
the paramagnetic phase.

\section{Discussion}

We have considered a Heisenberg model that includes isotropic exchange
interactions up to third nearest-neighbors, single-ion anisotropy and
long-range magnetic dipole-dipole interactions to describe the
long-range ordered state of the Gd$_2$Sn$_2$O$_7$ pyrochlore antiferromagnet.
The ground state of this system, as found by neutron scattering experiments,
corresponds to the (classical, PC) 
ground state described by Palmer and Chalker \cite{PC}
for the classical Heisenberg pyrochlore  with nearest-neighbor antiferromagnetic
exchange and long-range dipolar couplings~\cite{Raju:GTO,Cepas:MC,Cepas:MFT}.
We used a long wavelength ($1/\mathrm{S}$ spin wave) expansion to describe the low-energy
excitations about the PC ground state and to calculate the low-temperature
behavior of the specific heat, $C_v$, and order parameter, $m$, for this
material. 

By fitting the available specific heat data in the low-temperature range (0.35~K $<T<$ 0.5~K),
we were able to procure an estimate of the exchange interactions beyond nearest neighbors. 
We obtained evidence that Gd$_2$Sn$_2$O$_7$ is in a region of exchange coupling with 
large stability against quantum fluctuations.
Our main result (which \emph{does not} 
rely on excruciatingly fine-tuned exchange constants beyond
nearest-neighbor) is that the experimental temperature range 
0.35 K $<T<$ 0.5 K corresponds to the upper temperature range below which 
the thermodynamic quantities become thermally activated above an
excitation gap $\Delta \sim 1$~K \cite{DelMaestro:GTO}.
In other words, the  independently experimentally determined microscopic 
nearest-neighbor exchange (on the basis of DC magnetic susceptibility), 
single-ion anisotropy (on the basis of ESR) and dipolar coupling strength 
already predict a temperature dependence for $C_v$ in 
Gd$_2$Sn$_2$O$_7$ that is in rough agreement with the experiment without
significant adjustment. 

The excitation gap takes its origin from the combination of single-ion anisotropy  and
magnetic dipolar anisotropy.
From our fits of the experimental specific heat, we tentatively conclude that
the real gap is actually even \textit{larger} than the one we have determined.
Specifically, considering the lower temperature range in Fig.~\ref{fig:logCvvs1oT} and 
Fig.~\ref{fig:CvvsT} (and the inset of Fig.~\ref{fig:CvvsT} for $m$), it appears that 
the specific heat is dropping faster in the lower temperature range than the calculations predict.
We speculate that this may indicate that the sub-leading anisotropy terms 
neglected in ${\cal H}_{\rm cf}$ in Eq.~(5) (and which correspond to crystal field
terms $B_{l,m}$ with $l=4,6$) would further increase the effective gap.
In particular, those corrections would resign to further limit the
spin fluctuations perpendicular to the local three-fold axis. 
However, at this time, experimental measurements of $C_v$ below 0.3~K are required
to ascertain quantitatively the detail of the microscopic parameters for
Gd$_2$Sn$_2$O$_7$ and to determine with better precision the exchange parameters
$J_1$, $J_2$ and $J_{31}$.
As in other Gd$^{3+}$--based insulating magnetic materials~\cite{Cone-1,Cone-2},
it is also possible that anisotropic exchange interactions ultimately need
to be included in a complete description of Gd$_2$Sn$_2$O$_7$.

It therefore appears that a rejoinder to the question posed in the Introduction is that 
the low temperature specific heat observed in gadolinium stanate 
(Gd$_2$Sn$_2$O$_7$)
may possibly be well described by the 
conventional gapped spin wave excitations of Ref.~[\onlinecite{DelMaestro:GTO}].
We believe that either a confirmation or rebuttal of our
suggestion of gapped excitations in Gd$_2$Sn$_2$O$_7$ via 
specific heat ($C_v$) measurements down to $\sim 100$ mK 
would tremendously help focus the discussion about the pervasive
low energy excitations in insulating magnetic rare-earth pyrochlore oxides.
However, a possible confirmation of such gapped excitations in 
Gd$_2$Sn$_2$O$_7$ via specific heat measurements would ultimately
have to be rationalized within the context of the perplexing and persistent 
temperature-independent spin dynamics found in muon spin relaxation studies on this, and
other geometrically frustrated pyrochlores.

\section{Acknowledgments}

We are indebted to Pierre Bonville for kindly
providing us with the low temperature specific heat and 
M\"ossbauer data used for all the fits reported in this study.  
We thank 
Matt Enjalran, Tom Fennell and Mike Zhitomirsky for 
useful and stimulating discussions.
Support for this work was provided by the NSERC
of Canada and the Canada Research Chair Program (Tier I) (M.G),
the NSERC of Canada Grant PGS D2-316308-2005 (A.D.),
the Canada Foundation for Innovation,
the Ontario Innovation Trust, and the Canadian 
Institute for Advanced research.
M.G. acknowledges the University of Canterbury for an Erskine Fellowship
and the hospitality of
the Department of Physics and Astronomy at the University of Canterbury
where part of this work was completed.


\bibliography{gd2sn2o7-excite}

\end{document}